\documentclass[aps,pra,twocolumn,showkeys,nofootinbib,superscriptaddress]{revtex4}
\pdfoutput=1

\usepackage[utf8]{inputenc}
\usepackage{graphicx}
\usepackage{graphics}
\usepackage{epsfig}
\usepackage{lscape}
\usepackage{color}
\usepackage{multirow}
\usepackage{hyperref}
\usepackage{breqn}
\usepackage{physics}
\usepackage{subfiles}

\newcommand{\Chemie}{
Department Chemie,
Johannes Gutenberg-Universit\"at, Fritz-Strassmann Weg 2, 55128 Mainz, Germany \\
}
\newcommand{\HIM}{
Helmholtz-Institut Mainz,
Staudingerweg 18, 55128 Mainz, Germany \\
}
\newcommand{\RUG}{
Van Swinderen Institute for Particle Physics and Gravity,
University of Groningen, Nijenborgh 4, 9747 Groningen, The Netherlands\\
}
\newcommand{\GSI}{
GSI Helmholtzzentrum f\"ur Schwerionenforschung,
Planckstrasse 1, 64291 Darmstadt, Germany\\
}
\newcommand{\US}{
Science Department, Chatham University, Pittsburgh, Pennsylvania 15232, USA\\
}

\begin{document}
\title{State-specific ion mobilities of Lr$^+$ (Z = 103) in helium}
\author{Harry Ramanantoanina\footnote{Corresponding author: harry.ramanantoanina@kit.edu}}
\affiliation{\Chemie} \affiliation{\HIM}
\author{Anastasia Borschevsky}
\affiliation{\RUG}
\author{Michael Block}
\affiliation{\Chemie} \affiliation{\HIM} \affiliation{\GSI} 
\author{Larry Viehland}
\affiliation{\US}  
\author{Mustapha Laatiaoui\footnote{Corresponding author: mlaatiao@uni-mainz.de}}
\affiliation{\Chemie} \affiliation{\HIM}
\date{\today}

\begin{abstract}
Ion mobilities of Lr$^+$ ($Z = 103$) and of its lighter chemical homolog Lu$^+$ ($Z = 71$) in helium were calculated for the ground state $^1$S$_0$ and the lowest metastable state $^3$D$_1$. To this end we applied the multi-reference configuration interaction (MRCI) method to calculate the ion-atom interaction potentials in the different states. The Gram-Charlier approach to solving the Boltzmann equation was used to deduce the mobilities of the different electronic states, based on the calculated interaction potentials. We found that the zero-field ion mobilities are similar for the Lr$^+$ and Lu$^+$ ions. In addition, the ion mobilities of the different states are substantially different for temperatures above $100\,$K. The relative differences between the mobilities of the ground and excited states at room temperature are about $15$\% and $13$\% for Lu$^+$ and Lr$^+$ ions, respectively, which should be sufficiently large enough to enable laser resonance chromatography (LRC) of these ions.
\end{abstract}
\maketitle

\section{Introduction}
Lawrencium is a synthetic element that was first discovered in 1961 by A. Ghiorso and colleagues and has since been placed 103rd in the periodic table as the heaviest actinide \cite{Ghiorso:1961}. Interest in this element has not waned since then, motivated initially by new insights into the nuclear structure of its isotopes \cite{Asai:2015}, and not least by the question of whether lawrencium, together with lutetium, are  homologues of scandium and yttrium \cite{Jensen:1982,Scerri:2009}. 
Recently, it has gained special attention because its ionization potential has been studied for the first time \cite{Sato:2015} and laser spectroscopy has already reached its neighboring element nobelium \cite{Laatiaoui:2016,Chhetri:2018}, element 102, which in turn raises hopes to also experimentally study the atomic structure of lawrencium \cite{Laatiaoui:2019}. The dramatic increase of relativistic effects with atomic number makes the ground state of this element adopt the $7s^27p_{1/2}$ ($^2$P°$_{1/2}$) electron configuration, in contrast to lutetium, where the valence electron occupies the $d$ orbital. Various theoretical predictions agree with high confidence on this ground-state configuration in Lr \cite{Borschevsky:2007b, Fritzsche:2007,Dzuba:2014,Asai:2015,KahRaeEli21,Xu2016}, while experimental confirmation is still awaited.\par
In preparation for experiments, much theoretical work has been done in recent years to elucidate the atomic structure and ionization potential of neutral Lr \cite{Borschevsky:2007b,Fritzsche:2007,Dzuba:2014, KahRaeEli21}, taking into account relativistic effects (including, in many cases, the quantum electrodynamics corrections) and adequately addressing electron correlation. With the recent proposal to perform spectroscopy on superheavy ions using Laser Resonance Chromatography (LRC) techniques \cite{Laatiaoui:2020a}, the electronic structure of the singly charged ion has also come increasingly into focus \cite{Paez:2016,Kahl:2019,Lackenby:2020}.\par
In the LRC technique, the ions are subjected to pulsed laser beams for resonant optical pumping into metastable states before their release into a drift tube filled with helium gas. When the ions in different electronic states experience different interactions with helium atoms they move under the influence of an external and homogeneous electric field with different velocities through the drift tube toward the particle detector, enabling state-specific ion separation and resonance detection \cite{Laatiaoui:2020a}.  The transport properties of several actinide ions in noble gases have been studied already in Ref. \cite{Visentin:2020}, demonstrating the dependence of the ion mobility on the electronic configuration. In order to take advantage of the novel LRC approach, experimental conditions must be found such that the mobility of the ground state is substantially different from that of the excited state, which requires rigorous theoretical parameter confinement in advance \cite{Laatiaoui:2020b}. So far, only the ground-state transport properties have been predicted for the Lr$^+$ ion, and the excited-state mobilities remain to be explored for optimal design of the experiment.\par
In previous work \cite{Ramanantoanina2021,Ramanantoanina2022}, the multi-reference configuration interaction (MRCI) method provided reliable predictions of the energy levels for the ground and low-lying excited states of heavy metal ions, achieving theoretical uncertainties between $5$ and $10\,$\%.  In the current study we use the MRCI method to treat interatomic interactions; based on the obtained interaction potentials, we calculate the state-specific mobility of Lu$^+$ and Lr$^+$ ions drifting in helium gas in their ground ($^1$S$_0$) and lowest excited (metastable $^3$D$_1$) electronic states.

\section{Methodology and Computational Details}
The \textit{ab initio} MRCI calculations of the interaction potentials ($V(d)$) were performed using the DIRAC19 code \cite{DIRAC19}. The calculations were carried out in the framework  of the four component Dirac-Coulomb Hamiltonian, and the nuclei were treated within a finite-nucleus model \textit{via} the Gaussian charge distribution \cite{VisDya97}. The uncontracted Gaussian-type Dyall basis sets \cite{dyall2004,dyall2011} of single-augmented triple-zeta (s-aug-cv3z) quality were used for all the elements. The metal ion and the neutral helium atom were placed along the \textit{z}-axis in a system of Cartesian coordinates, separated by an inter-atomic distance $d$ that was varied from $2.0\,$\AA~ to $40.0\,$\AA~ for the calculation of the interaction potentials. We use the Boys-Bernardi counterpoise correction to tackle basis set superposition error \cite{Boys1970}: $V(d)=E_{M^+-He}(d)-E_{M^+}(d)-E_{He}(d)$, with M = Lu and Lr. $E_{M^+-He}(d)$ is the MRCI energy of the M$^+$-He system at an inter-atomic distance $d$. $E_{M^+}(d)$ and $E_{He}(d)$ are the energies of the systems M$^+$-Gh and Gh-He, respectively, where He and M atoms are replaced by a ghost atom (Gh) without charge but carrying the the full basis sets of the He and M elements, respectively.
\par
The electronic structure was obtained in two steps. In the first step, Dirac-Hartree-Fock calculations were performed using the average of configuration (AOC) type calculation. The AOC allowed us to represent the open-shell electronic structure system with 2 valence electrons that were evenly distributed over 12 valence spinors (6 Kramers pairs) of \textit{s} and \textit{d} atomic characters. The resulting wavefunction was used as reference for the CI calculations.  In the second step, the energy levels and the spectroscopic properties were calculated using the MRCI approach, within the Kramers-restricted configuration interaction (KRCI) module in DIRAC19 \cite{DIRAC19,saue2020,thyssen2008,knecht2010}. In this implementation, the KRCI calculations use the concept of generalized active space (GAS) \cite{fleig2003}, which enables MRCI calculations with single and double electron excitations for different GAS set-ups \cite{saue2020}. The MRCI model \textit{a priori} takes into consideration the dynamical correlation of the active electrons \cite{fleig2012}. \par 

\begin{table}[htp]
\centering
\caption{Specification of the generalized active space (GAS) scheme used in the calculations of the Lu$^+$-He and Lr$^+$-He systems. See text for details.}
\label{table1}
\begin{tabular}{ l| c| c| c| c}
\hline\hline
GAS       & \multicolumn{2}{|c|}{Accumulated}  & Number of & Characters\footnote{For Lu$^+$-He and Lr$^+$-He, n = 6 and 7, respectively} \\
Space     & \multicolumn{2}{|c|}{Electrons}    & Kramers   &            \\
          & Min\footnote{\textit{m} and \textit{q} are variables that control the electron excitation process attributed to the selective GAS}           & Max                & pairs     &            \\
\hline
1         &  8-\textit{m} &  8  & 4       & (n-1)\textit{s}, (n-1)\textit{p} \\
2         & 24-\textit{q} & 24  & 8       & (n-2)\textit{f}, He 1\textit{s} \\
3         & 24            & 26  & 9       & n\textit{s}, (n-1)\textit{d}, n\textit{p} \\
4         & 26            & 26  & $\leq$ 30 a.u.     & Virtual \\
\hline\hline
\end{tabular}
\end{table}

We report in \autoref{table1} the GAS set-up together with the technical specifications that were important in the MRCI calculation. In total, we considered 4 GAS that were selectively chosen to activate 26 electrons within 21 semi-core and valence orbitals as well as virtual orbitals with energies below 30 atomic units, i.e. 194 and 199 for the Lu$^+$-He and Lr$^+$-He systems, respectively. Because the total number of configuration state functions was too large, we defined the parameters \textit{m} and \textit{q} to control the electron excitation process that occurred at the semi-core level. These parameters were set to \textit{m}=2 and \textit{q}=1, which signified that double- and single-electron excitations were allowed from the selective GAS. It is noteworthy to point out that truncated configuration interaction method is not size-consistent \cite{Szalay2012}. We did not explicitly use the Davidson (+Q) corrections \cite{DIRAC19} to solve this problem. But we surmise that including higher order excitation in the GAS scheme (see \autoref{table1}) has helped to mitigate the size-concistency issue in the present MRCI method. Furthermore, we also employed size-extensive Fock space coupled cluster (FSCC) \cite{DIRAC19} to validate the MRCI method (\textit{vide infra}).
\par
To validate the MRCI results, we have also conducted calculations on the relativistic multireference FSCC level of theory. The relativistic FSCC approach is considered to be a very powerful method for the treatment of heavy atomic and small molecular systems and it is also available in DIRAC19 \cite{DIRAC19}; this method is particularly well-suited for treating systems with two valence electrons, via the sector(0,2) algorithm \cite{LanEliIsh04}, such as  Lu$^+$-He and Lr$^+$-He investigated here. \par
The FSCC calculation started with the closed-shell reference electronic state, that was, in our case, the Lr$^{3+}$-He and Lu$^{3+}$-He systems. For the sake of comparison, the FSCC computational details (relativistic method, basis sets, treatment of the nuclei) were the same as those used in the MRCI calculations (see above). In total, 60 and 74 core and semi-core electrons of the Lu$^+$ and Lr$^+$ ions, respectively, plus 2 He $1s$ electrons, were correlated. Virtual orbitals up to energies of 30 atomic units were also included in the correlation space. Then, two electrons were added to the selected virtual orbitals (the model space) to obtain the singly ionized Lr$^{+}$-He and Lu$^{+}$-He systems, for which the appropriate coupled cluster equations were solved in an iterative way. Convergence difficulties were lifted by complementing the FSCC method with the intermediate Hamiltonian approach \cite{LanEliIsh04}. \par
We described the electronic states that corresponded to the interaction potential between the heavy metal ions and the neutral He atom by means of the quantum numbers \textit{J} and $\Omega$, representing the total angular momentum of the metal ions and its projection onto the inter-atomic axis, respectively. In particular, we labelled the electronic states using the Hund’s case (c) notation, i.e. $\Omega_J^\sigma$, for consistency with conventional practice for the calculations of ion mobility and transport properties ($\sigma$ = $+$ or $-$ is an additional notation proper to the linear $C_{\infty v}$ point group that represents the invariability of the wavefunction with respect to the $\sigma_v$ symmetry operator) \cite{Laatiaoui:2020b,Visentin:2020}. Thus, the ground state $^1$S$_0$ of the free Lr$^+$ and Lu$^+$ ions give rise to the $\Omega$ = 0 state in the Lr$^+$-He and Lu$^+$-He systems, i.e. $X0^+$. The metastable $^3$D$_1$ state, on the other hand, transformed to the non-degenerate $0_1^-$ and double-degenerate $1_1$ states,  since $\Omega$ = 0 and $\pm$1, respectively. Using the same notation, the next excited states $^3$D$_2$ and $^3$D$_3$ transformed to the single-degenerate $0_2^+$ and $0_3^-$ as well as the double-degenerate $1_2$, $1_3$, $2_2$, $2_3$, and $3_3$ states.\par
The ion mobilities were calculated from the ion-atom interaction potentials by solving the Boltzmann equation by the Gram-Charlier approach~\cite{Viehland:2018}. To this end we used the program PC~\cite{Viehland:2010}, which delivers the momentum transfer and other transport cross sections as a function of the collision energy. From this we then calculated the reduced ion mobility $K_0$ either as a function of temperature at a given electric field-to-gas-number density ($E/n_0$) or as a function of $E/n_0$ at different temperatures, utilizing the program VARY~\cite{Viehland:2012}. Here $K_0$ is the ion mobility $K$ normalized to the standard pressure $P_0$ and the standard temperature $T_0$ according to $K_0=K\frac{P}{P_0}\frac{T_0}{T}$. Beyond $d=40\,$\AA, the interaction potentials were adjusted to asymptotically mimic the long-range induced ion-dipole attraction given by $V_{pol}(d)=e^2 \alpha_p/(2(4\pi\epsilon_0)^2 d^4)$, with the static average dipole polarizability of helium of $\alpha_p=0.205\,$\AA$^3$~\cite{Lide:1989}. 

For the $^1$S$_0$ ground state, there is only one potential per ionic species and thus a single mobility curve was obtained for each ion.
For the $^3$D$_1$ excited state an isotropic averaged potential was first used to calculate the isotropic ion mobility $K_0^{iso}$, which we then compared with the averaged mobility $K_0^{av}$~\cite{Visentin:2020}. This latter was obtained by averaging the mobilities from interaction potentials for each $\Omega$ component with their statistical weights according to
\begin{dmath}
K_0^{av}(T)=[K_0^{0_1^-}(T)+2K_0^{1_1}(T)]/3.
\label{eq1}
\end{dmath}
In the Supplementary Material, Figure S1 shows the calculated isotropic ion mobilities together with the averaged ion mobilities corresponding to the $^3$D$_1$ state for both Lu$^+$-He and Lr$^+$-He systems.

\section{Results and Discussion}

\autoref{table2} lists the calculated energies of the ground ($^1$S) and the metastable ($^3$D) electronic states of the Lu$^+$ and Lr$^+$ obtained from the MRCI and the FSCC calculations. To calculate these energies following the models described in the Methodology section, we set the distance between the metal ion and the neutral atom to 40.0 Å, which is large enough to make sure that the energy difference between the multiple components of the free ion multiplets are negligible, so that they are comparable to atomic energies. For Lu$^+$, experimental data are also listed for comparison. The two theoretical models at hand yield similar results, with good agreement with the experiments and previous theoretical data \cite{Kahl:2019,Ramanantoanina2022}. The good agreement with both the experimental values and the FSCC results confirms the suitability of MRCI for electronic structure calculations of heavy metal ions. \par

\begin{table}[htp]
\centering
\caption{Calculated low-lying energy levels of Lu$^+$-He and Lr$^+$-He ions (in cm$^{-1}$) obtained from the FSCC and MRCI calculations (the separation distance between the metal Lu$^+$/Lr$^+$ ions and He are set to 40.0 Å, showing the energy of the ground state $^1$S and metastable states $^3$D of the metal ions.). For Lu$^+$-He, the experimental energy values reported for the Lu$^+$ ion is also listed for comparison.}
\label{table2}
\begin{tabular}{ l| c| c| c| c| c| c}
\hline\hline
                 &            &\multicolumn{3}{c|}{Lu$^+$}&\multicolumn{2}{c}{Lr$^+$}\\
$^{2S+1}L_J$&$\Omega_J^\sigma$&FSCC &MRCI &Exp.\footnote{taken from ref. \cite{NISTdata}}            &FSCC&MRCI\\
\hline
$^1$S$_0$   &$X0^+$                   &    0&    0&    0&    0&    0\\
$^3$D$_1$   &$0_1^-$+$1_1$            &12354&12184&11796&20265&21751\\
$^3$D$_2$   &$0_2^+$+$1_2$+$2_2$      &12985&12642&12435&21623&22442\\
$^3$D$_3$   &$0_3^-$+$1_3$+$2_3$+$3_3$&14702&13881&14199&26210&24708\\
\hline\hline
\end{tabular}
\end{table}

\begin{figure*}
\includegraphics[width=1.00\textwidth]{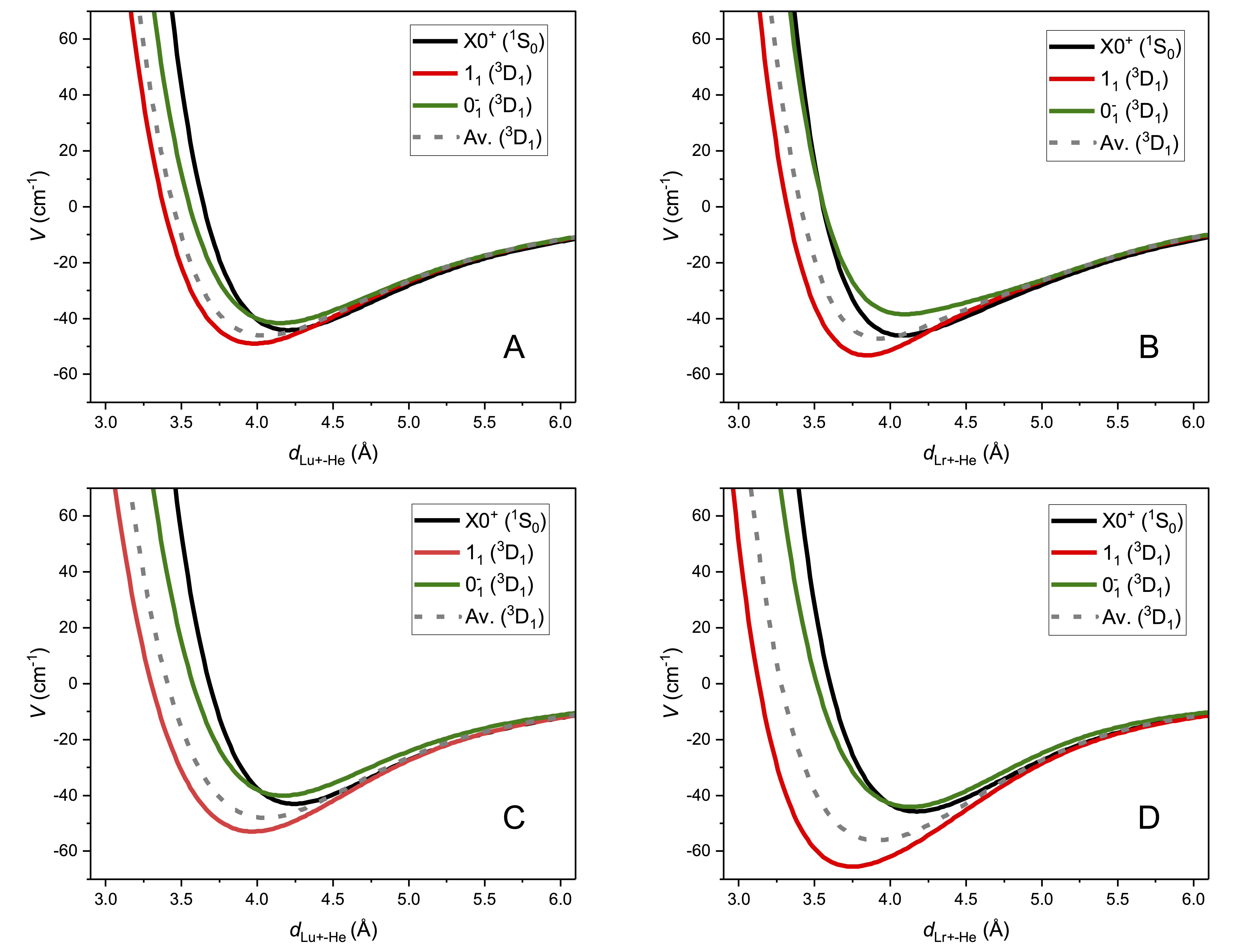}
\caption{Graphical representations of the MRCI interaction potentials of the Lu$^+$-He (A) and Lr$^+$-He (B) systems, compared with the FSCC interaction potentials of Lu$^+$-He (C) and Lr$^+$-He (D). In each panel, the ground XO$^+$ ($^1$S$_0$) (solid black curve), and the two low-lying metastable $1_1$ ($^3$D$_1$) (red) $0_1^-$ ($^3$D$_1$) (green) states are represented, together with the calculated average interaction potential for the metastable states (dashed grey curve). Note that for clarity the potentials are normalized to the same dissociation limit.}
\label{figure1}
\end{figure*}

The calculated interaction potentials of the ground and the metastable electronic states of Lu$^+$-He and Lr$^+$-He are shown in \autoref{figure1}. Since the MRCI and FSCC calculations are both based on the relativistic Dirac-Coulomb Hamiltonian and employ the same basis sets, the discrepancies between the two sets of calculations can be interpreted in terms of the treatment of electron correlation. There is a good agreement between the FSCC and MRCI data of the ground-state X0$^+$ ($^1$S$_0$) for both Lr$^+$-He and Lu$^+$-He systems. For the latter, the ground-state interaction potentials are also comparable with the scalar-relativistic potentials reported in Refs.\cite{Laatiaoui:2020b,Visentin:2020}. However, we obtain larger discrepancies between the two methods for the metastable states, namely in the $1_1$ ($^3$D$_1$).

The \textit{ab initio} interaction potentials are fitted by the Morse potential energy function \cite{Morse1929},
\begin{dmath}
V(d)=D_e(e^{-2\alpha(d-d_{min})}-2e^{-\alpha(d-d_{min})})
\label{eq2}
\end{dmath}
to derive the the equilibrium inter-atomic distance ($d_{min}$), the dissociation energy ($D_e$), and the range parameter $\alpha$. These are listed in \autoref{table3} and \autoref{table4}. The energy splitting between the $1_1$ and $0_1^-$ ($^3$D$_1$) metastable states in the short-range interaction is larger in the FSCC data. Thus we see a slight shift of the equilibrium distances ($d_{min}$) to smaller values from the MRCI to the FSCC $1_1$ ($^3$D$_1$) interaction potentials: 0.06 $\AA$  and 0.12 $\AA$ for the Lu$^+$-He and Lr$^+$-He systems, respectively. We also see a slightly higher FSCC dissociation energy of this state in Lr$^+$-He. \par

The calculated $d_{min}$ values by the two models follow the same trend in the Lu$^+$-He (\autoref{table3}) and Lr$^+$-He (\autoref{table4}) systems: $d_{min}$ associated with the X0$^+$ ($^1$S$_0$) ground state is larger compared with the two $1_1$ and $0_1^-$ ($^3$D$_1$) metastable states. We can compare the spectroscopic parameters for Lu$^+$-He in both the ground and metastabe states (cf. \autoref{table3}) as well as Lr$^+$-He in the ground state (cf. \autoref{table4}) to earlier coupled cluster studies \cite{Laatiaoui:2020b,Visentin:2020}. We find that the present $d_{min}$ values (both MRCI and the FSCC values) are slightly larger ($+2.5$\%) than the results reported in the aforementionned Refs. \cite{Laatiaoui:2020b,Visentin:2020} for the ground and the metastable states. \par
The dissociation energy ($D_e$) calculated within the MRCI and the FSCC models also follows a similar trend in the two systems. The metastable  $0_1^-$ ($^3$D$_1$) shows the weakest interaction, whereas its counterpart $1_1$ ($^3$D$_1$) state has the strongest. This is in contrast to the earlier predictions (\autoref{table3}), where the calculated $D_e$ for the $0_1^-$ and $1_1$ ($^3$D$_1$) metastable states in Lu$^+$-He are very close (with an energy difference of 2.3 cm$^{-1}$ only). This energy difference is 6.94 cm$^{-1}$ and 13.1 cm$^{-1}$, respectively, for the present MRCI and FSCC results (\autoref{table3}). This disagreement with previous calculations could be due to the difference in the treatment of relativistic effects. We used the four-component DCHF-based model as a basis of our calculations, whereas in Ref. \cite{Laatiaoui:2020b}, the spin-orbit coupling is treated within perturbation theory based on a scalar relativistic electronic structure. However, the many different computational parameters (choice of basis set and the correlation space and the treatment of correlation) make direct comparison between these values difficult.
Overall, the bonding interaction between the metal ion and the helium atom is shown to be very weak, regardless of the theoretical approach.  
\begin{table}[htp]
\centering
\caption{Calculated spectroscopic dissociation energy $D_e$ (in cm$^{-1}$), equilibrium distance $d_{min}$ (in \AA), and range parameters $\alpha$ (in 1/\AA) of the interaction potential of Lu$^+$-He corresponding to the Lu$^+$ $^1$S$_0$ and $^3$D$_1$ electronic states obtained by a mathematical fit to Equation~\ref{eq2}, and compared with previous calculations (Ref.).}
\label{table3}
\begin{tabular}{ l| c| c| c| c| c| c| c| c| c}
\hline\hline
\multicolumn{2}{c|}{Lu$^+$-He}&\multicolumn{3}{c|}{MRCI}&\multicolumn{3}{c|}{FSCC}&\multicolumn{2}{c}{Ref.\footnote{taken from ref. \cite{Laatiaoui:2020b}}} \\
\multicolumn{2}{c|}{}         &$D_e$ &$d_{min}$&$\alpha$&$D_e$ &$d_{min}$&$\alpha$&$D_e$ &$d_{min}$\\
\hline          
$^1$S$_0$ & X0$^+$            &45.23 &4.224    &1.213   &42.71 &4.264    &1.197   &47.3  &4.17\\
$^3$D$_1$ & 0$_1^-$           &42.13 &4.171    &1.133   &39.16 &4.187    &1.123   &49.9  &4.11\\
          & 1$_1$             &49.07 &4.006    &1.114   &52.26 &3.948    &1.075   &52.2  &3.91\\
\hline\hline
\end{tabular}

\end{table}

\begin{table}[htp]
\centering
\caption{Calculated spectroscopic dissociation energy $D_e$ (in cm$^{-1}$), equilibrium distance $d_{min}$ (in \AA), and range parameters $\alpha$ (in 1/\AA) of the interaction potential of Lr$^+$-He corresponding to the Lr$^+$ $^1$S$_0$ and $^3$D$_1$ electronic states, obtained by a mathematical fit to Equation~\ref{eq2}, and compared with previous calculations (Ref.).}
\label{table4}
\begin{tabular}{ l| c| c| c| c| c| c| c| c| c}
\hline\hline
\multicolumn{2}{c|}{Lr$^+$-He}&\multicolumn{3}{c|}{MRCI}&\multicolumn{3}{c|}{FSCC}&\multicolumn{2}{c}{Ref.\footnote{taken from ref. \cite{Visentin:2020}}} \\
\multicolumn{2}{c|}{}         &$D_e$ &$d_{min}$&$\alpha$&$D_e$ &$d_{min}$&$\alpha$&$D_e$ &$d_{min}$\\
\hline          
$^1$S$_0$ & X0$^+$            &46.78 &4.134    &1.192   &48.59 &4.179    &1.205   &52    &4.08\\
$^3$D$_1$ & 0$_1^-$           &39.49 &4.177    &1.118   &43.92 &4.108    &1.174   &      &\\
          & 1$_1$             &51.84 &3.880    &1.094   &72.37 &3.756    &1.111   &      &\\
\hline\hline
\end{tabular}
\end{table}

\begin{figure}
\includegraphics[width=0.45\textwidth]{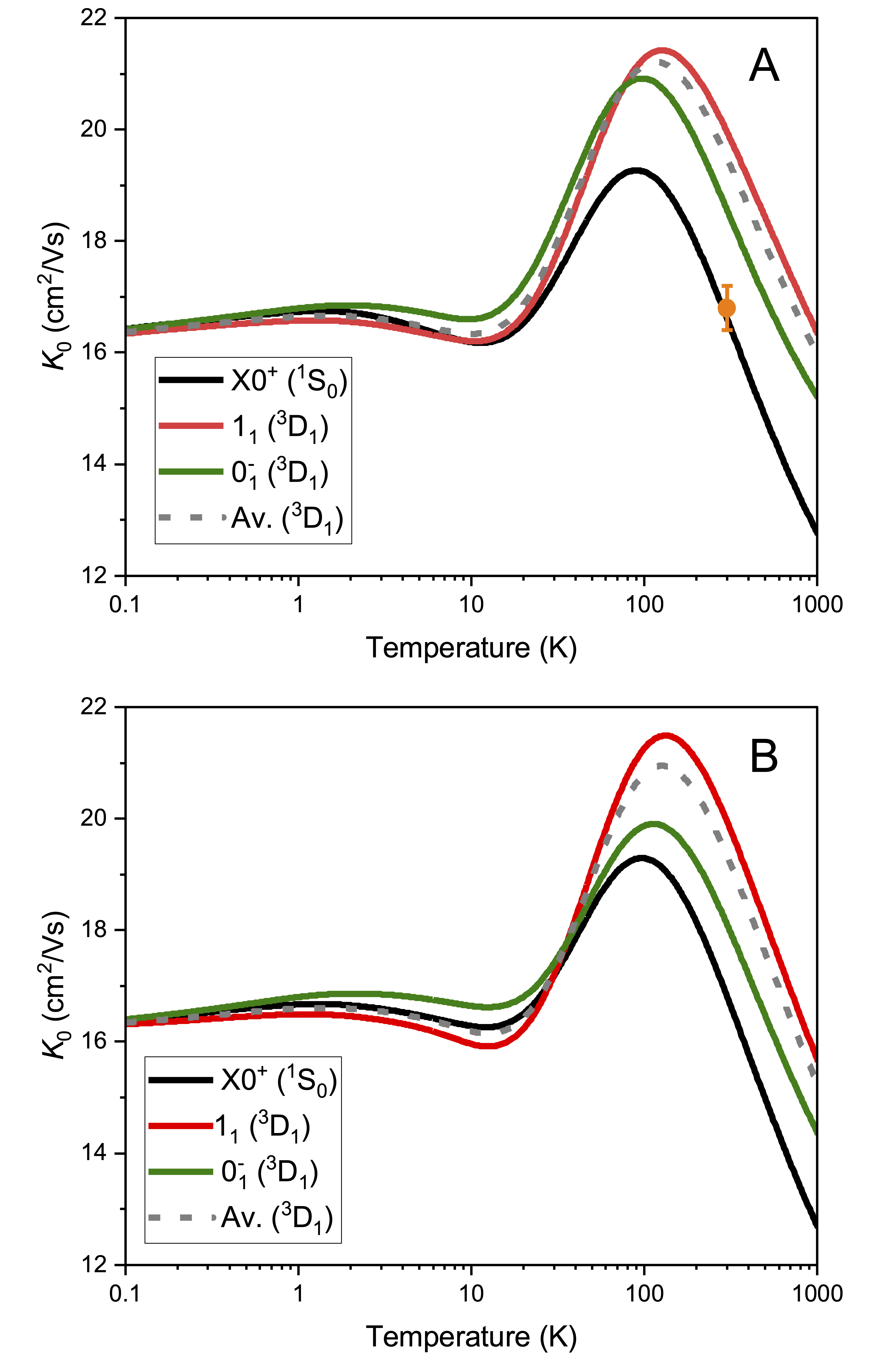}
\caption{Reduced zero-field mobilities of the Lu$^+$-He (A) and Lr$^+$-He (B) systems in the ground X0$^+$ ($^1S_0$) (in black), as well as in the metastable $1_1$ ($^3$D$_1$) (in red) and the $0_1^-$ ($^3$D$_1$) (in green)  states as a function of the temperature, derived from the MRCI interaction potential. The calculated average mobilities for the metastable states ($^3$D$_1$) are also depicted (in grey); the data point that represents the experimental result for the ground state of Lu$^+$ ion is also shown (orange dot and error bar)~\cite{Manard:2017}.}
\label{figure2}
\end{figure}

\begin{figure}
\includegraphics[width=0.45\textwidth]{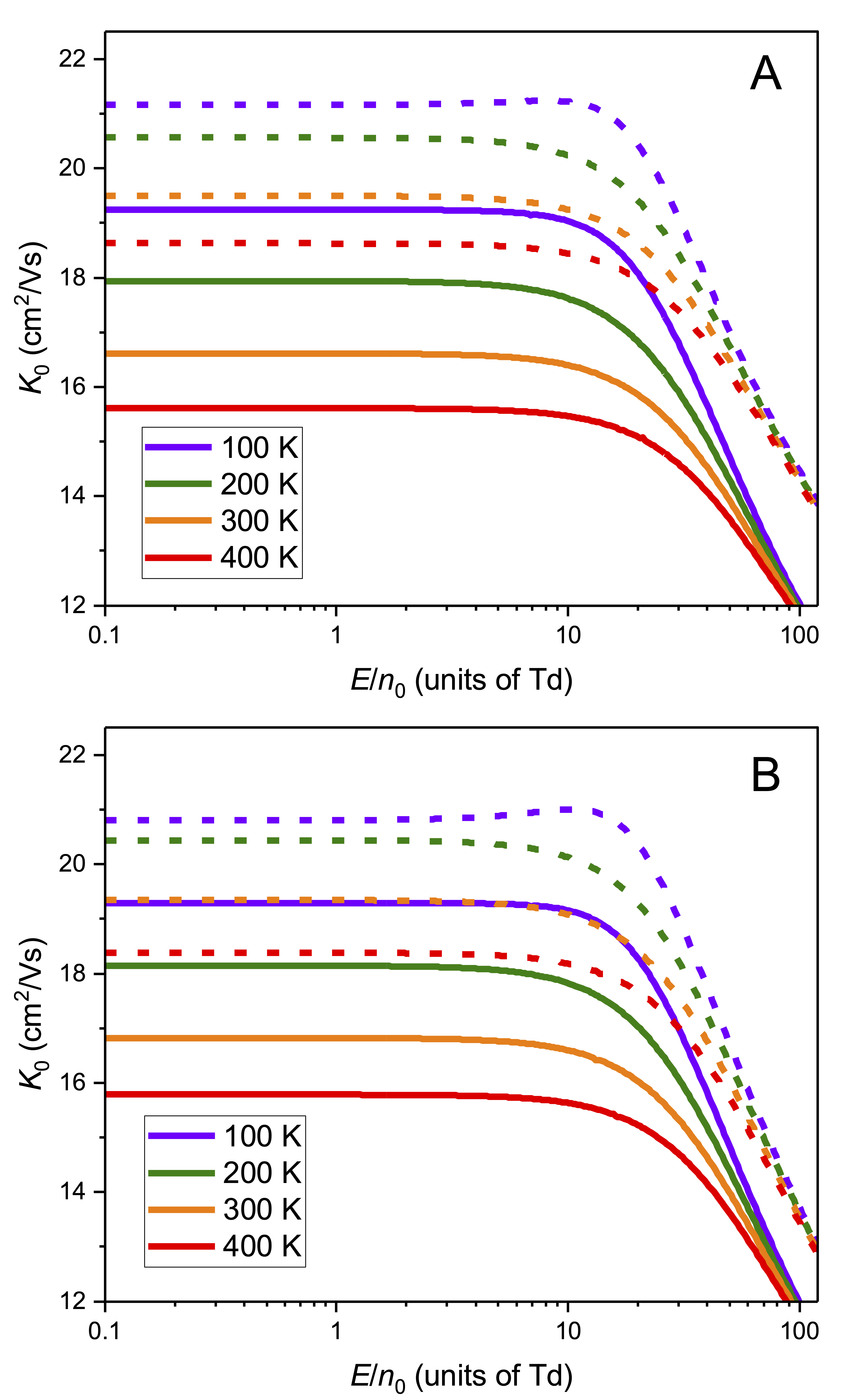}
\caption{Reduced mobilities of the Lu$^+$-He (A) and Lr$^+$-He (B) systems as function of E/n$_0$ and at selected temperatures, derived from the MRCI interaction potential, corresponding to the ground X0$^+$ ($^1S_0$) (solid lines) and the metastable $^3$D$_1$ (dashed lines) states. For the metastable states, the depicted mobility curves correspond to the average calculated mobility of the $1_1$ and the $0_1^-$ electronic states obtained using Equation~\ref{eq1} (See also the Supplementary Material, Figure S4).}
\label{figure3}
\end{figure}

The ion mobilities were calculated based on the interaction potentials obtained from the MRCI calculations. Figure \ref{figure2} shows an overview of the obtained zero-field mobilities of Lu$^+$ and Lr$^+$ in helium as function of the gas temperature. We have also calculated the ion mobilities based on the FSCC interaction potentials. In the Supplementary Material (Figure S2), a comparison between mobilities obtained with the MRCI and FSCC interaction potentials are also presented. In general, the mobilities of the ground states of the two ions obtained based on the different \textit{ab initio} approaches agree well with each other. In addition a good agreement of the Lu$^+$-He and Lr$^+$-He ground-state mobilities is achieved with those reported in Refs. \cite{Laatiaoui:2020b,Visentin:2020} \par
The principal difference between the two method concerns the $\Omega=1$ components; the MRCI calculations result in smaller dissociation energies at larger equilibrium distances (see Table ~\ref{table3} and Table ~\ref{table4}) and thus in smaller state-specific reduced mobilities compared with the FSCC based predictions, see also the Supplementary Material Figure S2. For brevity we discuss here only the mobility results obtained from predictions based on the MRCI interaction potentials; the minor differences between the MRCI and FSCC results mean that the conclusions of this work are not dependent on the computational approach. \par
The mobilities of the two systems are distinct for the different ionic states in a wide temperature range and converge towards the polarization limit $K_{pol}=(13.876/\alpha_p^{1/2})[(M_{He}+M_{ion})/M_{He}M_{ion}]^{1/2}$ at about $15.5\,$cm$^2$/Vs in the mK temperature regime~\cite{Viehland:1975}. Here, the number $13.876$ is obtained when $\alpha_p$ is given in units of \AA$^3$ and the masses $M$ are in atomic mass units. The mobility increases with increasing temperature to reach a local maximum at $100\,$K for the ions in the $^1$S$_0$ ground state (X0$^+$) before it decreases below $13\,$cm$^2$/Vs at temperatures beyond $1000\,$K. The predicted mobility for Lu$^+$ in the ground state is found to be in excellent agreement with the only available experimental data reported for a temperature of $295\,$K~\cite{Manard:2017}.\par
For the ions in the $^3$D$_1$ states we calculated the mobilities for each of the $\Omega$ components, 0$_1^-$ and 1$_1$, to deduce the average mobility $K_0^{av}$ according to Equation~\ref{eq1}. The resulting curves are indicated with dashed lines in Figure~\ref{figure2}. Although these average mobilities are deemed to be more accurate than the isotropic ones~\cite{Visentin:2020} we found that both mobilities are in excellent agreement with each other for the two ion species, Lu$^+$ and Lr$^+$ in the $^3$D$_1$ states, see the Supplementary Material Figure S1.
Similarly to the ground state mobilities, the average excited state mobility increases with increasing temperature to reach a maximum at about $105\,$K for both ions before it decreases towards higher temperatures. Noteworthy, however, is the difference in zero-field mobility between the ground state and the average mobility of the excited state: e.g., 14.8\% and 13\% at room temperature for the Lu$^+$-He and Lr$^+$-He systems, respectively, with a tendency of increasing relative differences towards higher temperatures. \par
In Figure ~\ref{figure3} we show the reduced mobilities of the different states of Lu$^+$ and Lr$^+$  obtained based on the MRCI interaction potentials for different gas temperatures as a function of the electric-field-to-gas-number density $E/n_0$ (the so-called reduced electric field, which is given in units of Townsend, $1\,$Td$=10^{-17}\,$Vcm$^2$). In the Supplementary Material, Figure S3 shows a comparison between the ion mobilities calculated with the MRCI and FSCC interaction potentials. In general, the mobility is roughly constant at reduced fields below $10\,$Td, such that it depends mainly on the gas temperature, with a tendency of being larger at lower temperatures (down to 100 K). As the reduced field increases, the ion mobility decreases almost exponentially to values below $12\,$cm$^2$/Vs at $E/n_0 \geq 100\,$Td for the ions in the ground states. It nearly decouples from the temperature dependency at extremely large $E/n_0$ values as the energy gained from the electric field dominates the effective ion temperature. The average mobility for the excited states exhibits a similar behavior as a function of the reduced field. Although the $0$-orbital-projection components have rather small mobilities, close to those of the ground states, the average mobilities of the excited states are dominated by the $\Omega=1$ component, due its higher multiplicity (Equation ~\ref{eq1}), see also the Supplementary Material Figure S4. Similar reduced mobilities were obtained for both investigated ionic species, Lu$^+$ and Lr$^+$, in the $^3$D$_1$ state, as can be seen in Figure ~\ref{figure3}.

\begin{figure}
\includegraphics[width=0.45\textwidth]{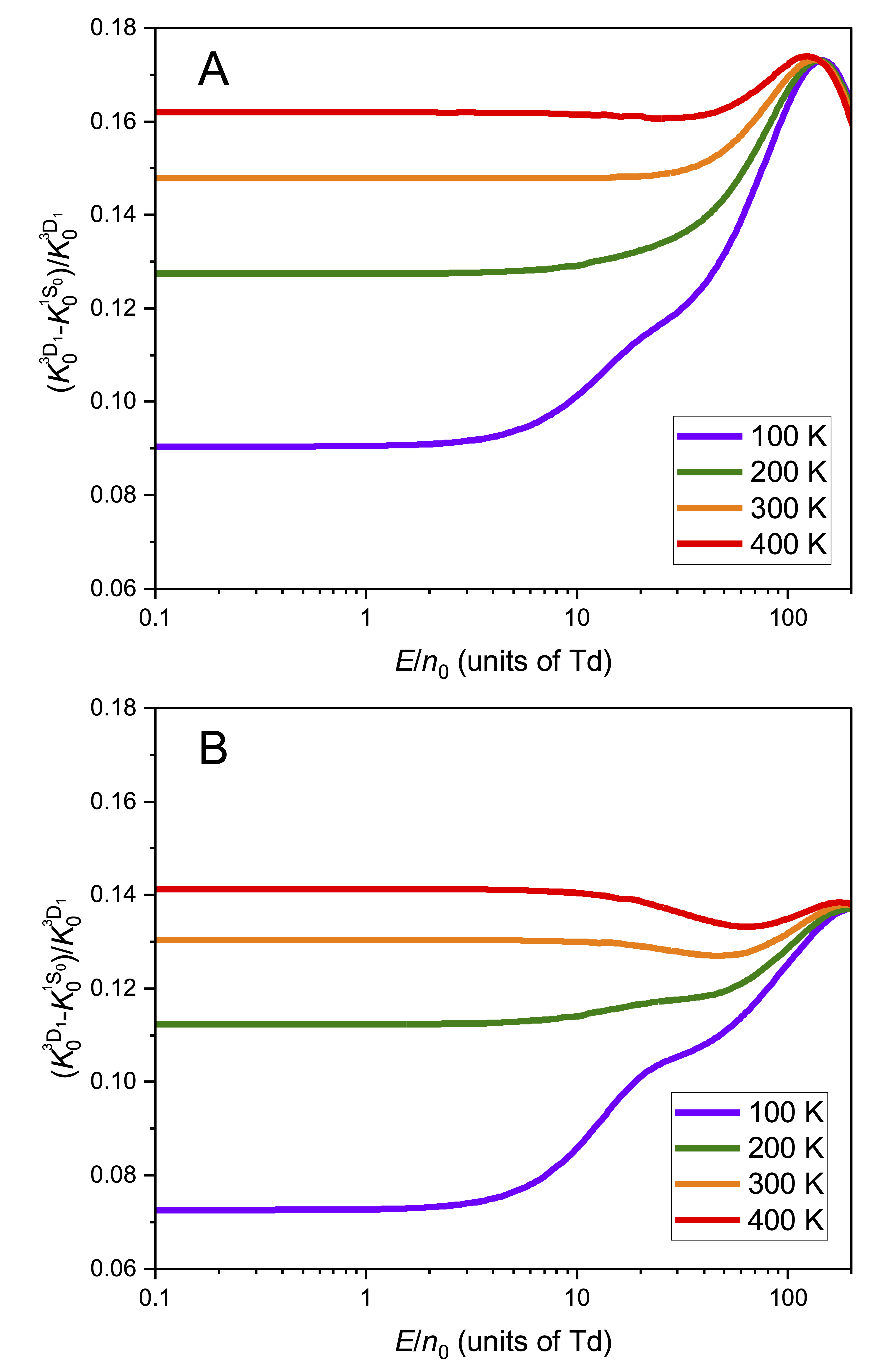}
\caption{Relative differences between the reduced mobilities of the ground and the average metastable states of the Lu$^+$-He (A) and Lr$^+$-He (B) systems as function of E/n$_0$ at selected temperatures, derived from the MRCI interaction potentials.}
\label{figure4}
\end{figure}

In order to reach significant time resolution in future LRC applications the relative drift time differences, due to the relative mobility differences, have to be maximized~\cite{Laatiaoui:2020b}.
Our results suggest a similar trend for both investigated ionic species, see Figure ~\ref{figure4}.
At gas temperatures above $100\,$K the relative mobility differences for the ground and excited states are between $7$\% and $17$\% at reduced fields below $200\,$Td, with a tendency of becoming larger for higher temperatures. These differences stay above $12$\% at room temperature and become rather insensitive to the reduced field at $400\,$K. 

\section{Summary and Conclusion}

We have carried out MRCI and FSCC calculations of the interaction potentials of the Lr$^{+}$-He and Lu$^{+}$-He systems. Based on these potentials we predicted the ion mobility of Lu$^+$ and Lr$^+$ ions in helium gas, and found the two approaches to be in excellent agreement, justifying the use of either method in future investigations. In particular, the MRCI-based method for calculating interaction potentials and ion mobilities will be relevant for the study of the Rf$^{+}$-He systems, which will be the next step of our theoretical work.\par
The predicted ion mobility value for Lu$^+$ in the ground state at room-temperature is in a striking agreement with the experimentally reported one ~\cite{Manard:2017b}. Similar accuracy can be expected for the room temperature mobility of Lr$^+$, at least in its ground state, due to the similarities of their electronic structures. As long as the reduced fields are below $200\,$Td, we expect the relative drift time differences to be above $7$\%; the higher the gas temperature the larger these differences become. Laser resonance chromatography on both ionic species should thus be feasible in terms of time resolution already at room temperature, i.e., without involving sophisticated cryogenic ion mobility spectrometers. Since quenching of states may become strong at elevated effective ion temperatures~\cite{Laatiaoui:2020b}, one may prefer to apply moderate reduced fields below $40\,$Td and room temperature gas environments in order to maintain state populations in the envisaged LRC experiments~\cite{RomeroRomero:2022}. In such a case we expect relative drift time differences between ground and metastable states to be about $15$\%  and $13\,$\% for Lu$^+$ and Lr$^+$, respectively, which should enable disentangling the different drift behaviours upon resonant excitations.

\section{Acknowledgements}
This project has received funding from the European Research Council (ERC) under the European Union’s Horizon 2020 Research and Innovation Programm (Grant Agreement No. 819957). We also gratefully acknowledge high performance computing (HPC) support, time and infrastructure provided by: SURFsara HPC at Snellius via the HPC-Europa3 programm, the Center for Information Technology of the University of Groningen (Peregrine), the Johannes Gutenberg University of Mainz (Mogon), and the HPC group of GSI. We thank Alexei Buchachenko for useful discussion.


\end{document}